\documentclass[10pt,twocolumn,twoside,english]{IEEEtran}
\usepackage{color}

\usepackage{cite}

\ifCLASSINFOpdf
   \usepackage[pdftex]{graphicx}
\else
  \usepackage[dvips]{graphicx}
   \DeclareGraphicsExtensions{.eps}
\fi
\usepackage{subfigure}

\usepackage{amsmath}
\usepackage{amsfonts,helvet}
\usepackage{amssymb}

\usepackage{algorithm}
\usepackage{algpseudocode}

\usepackage{array}
\usepackage{multirow}

\usepackage{hyperref}

\begin{document}
\title{Prototyping Real-Time Full Duplex Radios}


\author{MinKeun Chung,~\IEEEmembership{Student~Member,~IEEE}, Min Soo Sim,~\IEEEmembership{Student~Member,~IEEE}, Jaeweon~Kim,~\IEEEmembership{Member,~IEEE}, Dong Ku Kim,~\IEEEmembership{Member,~IEEE}, and Chan-Byoung~Chae,~\IEEEmembership{Senior~Member,~IEEE}\\
\thanks{M. Chung and M. S. Sim, D. K. Kim, and C.-B. Chae are with Yonsei University, Korea (E-mail: \{minkeun.chung, simms, dkkim, cbchae\}@yonsei.ac.kr). J. Kim is with National Instruments, TX, USA (E-mail: jaeweon.kim@ni.com). C.-B. Chae and D. Kim are co-corresponding authors.} 
\thanks{This work was in part supported by the the MISP under the ``IT Consilience Creative Program" (IITP- 2015-R0346-15-1008) and by the ICT R\&D Program of MSIP/IITP [B0126-15-1012].
 }}

\markboth{To appear in IEEE Communications Magazine}%
{Chung \MakeLowercase{\textit{et al.}}: Prototyping Real-Time Full Duplex Radios}

%

\maketitle
\begin{abstract}
In this article, we present a real-time full duplex radio system for 5G wireless networks. Full duplex radios are capable of opening new possibilities in contexts of high traffic demand where there are limited radio resources. A critical issue, however, to implementing full duplex radios, in real wireless environments, is being able to cancel self-interference. To overcome the self-interference challenge, we prototype our design on a software-defined radio (SDR) platform. This design combines a dual-polarization antenna-based analog part with a digital self-interference canceler that operates in real-time. Prototype test results confirm that the proposed full-duplex system achieves about 1.9 times higher throughput than a half-duplex system. This article concludes with a discussion of implementation challenges that remain for researchers seeking the most viable solution for full duplex communications.

\end{abstract}

\begin{IEEEkeywords}
Real-time full duplex radio, self-interference cancellation, software-defined radio, prototyping, dual-polarization antenna, 5th generation (5G) communications.
\end{IEEEkeywords}

%
\IEEEpeerreviewmaketitle


\section{Introduction}
\subsection{New Breakthrough: Full Duplex Radios}
How much does it cost to purchase a wireless spectrum? In a wireless spectrum auction in January of 2015, the Federal Communications Commission (FCC) raised, for a 65 MHz bandwidth, a record-breaking $\$$44.9 billion. This illustrates how valuable the wireless spectrum has become; it offers more high-speed connectivity and satisfies more user demand for data within a limited wireless spectrum. The FCC is considering releasing more spectrum for wireless broadband usage. For the endless surge in wireless data traffic, however, this cannot be the ultimate solution. 

Mobile devices with advanced wireless network capabilities, such as smartphones and tablets, are becoming ubiquitous, and keeping pace with their growth is the ever-increasing demand for bandwidth. Global mobile data traffic will increase nearly tenfold between 2014 and 2019. In that time, mobile data traffic is expected to grow at a compounded annual growth rate of 57~\%, reaching 24.3 exabytes per month by 2019 \cite{cisco2015}. These trends could create a {\it spectrum crunch} as the frequencies used to carry this traffic become exhausted. 

Although the laws of physics prohibit the production of more spectrum, there is a lot of potential for {\it aggressive expansion} in scarce resource, that is, boosting spectral efficiency using novel technologies. A candidate, for creating a new breakthrough to alleviate the {\it spectrum crunch}, is full duplex. It theoretically doubles spectral efficiency, making it worth billions of dollars. Full duplex thus holds the tremendous potential to carry out the solutions needed in the future evolution of wireless systems.

\subsection{Key Challenge: Self-interference} \label{Key Challenge}
Since Guglielmo Marconi developed the wireless telegraph in 1895, the bane of wireless networks has been self-interference. It is the presence of self-interference that represents the key challenge to implementing full duplex wireless systems. Self-interference is the phenomenon where, through the coupling of transceivers in a wireless network, a signal is transmitted from a transmitter to its own receiver while that receiver is attempting to receive a signal sent by the other device. It compels the fundamental assumption that a wireless network has to be operated in half-duplex mode on the same channel. For example, Long Term Evolution (LTE) frequency-division duplex (FDD) today is operated so that the downlink and uplink transmission take place in two different frequency bands. In other words, the existence of self-interference cuts in half the amount of resources available, such as time and frequency, for wireless communications. For this reason, it is essential to manage self-interference to achieve the highest throughput performance with limited radio resources.

\subsection{The Beginning of Aggressive Expansion: SDR Platform-based Prototyping} \label{Aggressive Expansion}
   
  \begin{figure*}
  \centering
  \includegraphics[width=6in]{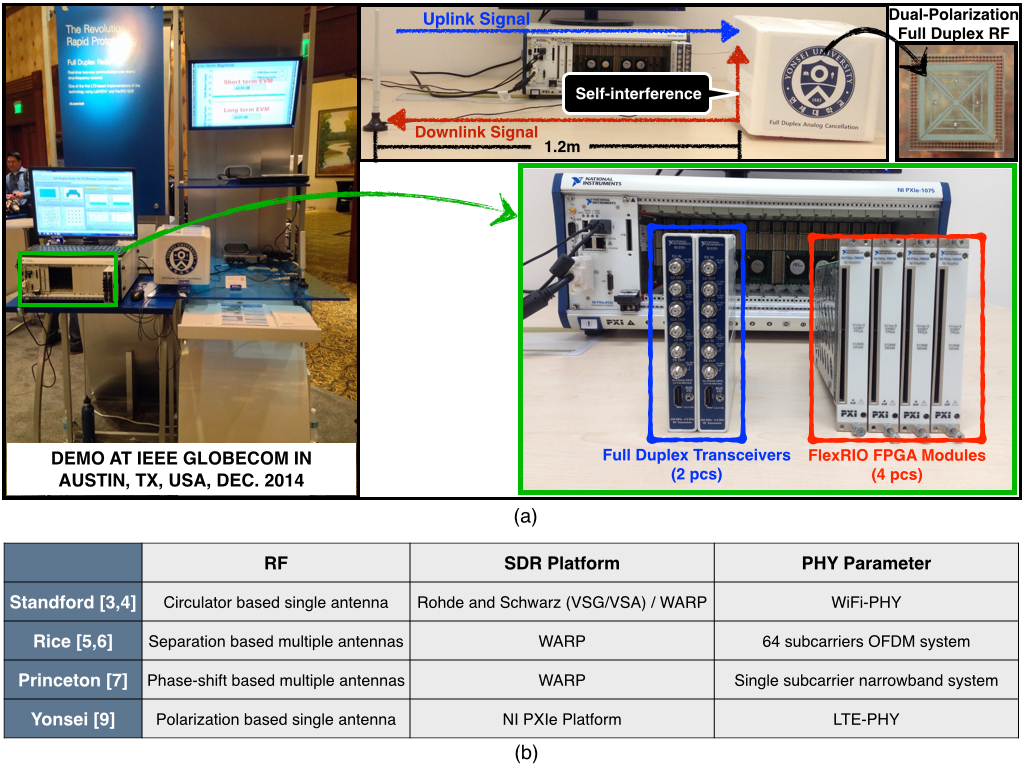}
  \caption{(a) Real-time full duplex radio experiment setup in our laboratory/the exhibition hall at IEEE Globecom in Austin, TX, USA, December 2014 \;(b) the categorized comparison of implementation characteristics by each research group. In the SDR platform, VSG and VSA denote vector signal generator and analyzer, respectively.} \label{fig:Exp_Setup}
  \end{figure*}
Up to this point, researchers have mostly depended on software simulations to test their theories that exploit simplified channel models (e.g., additive white Gaussian noise (AWGN), Rayleigh fading, etc.). In real-world wireless systems, however, impairments occur that are often overlooked in simulations, such as amplifier nonlinearity, gain/phase offset, I/Q imbalance, quantization effects, and timing jitter. Such impairments make prototyping imperative if the feasibility and commercial viability of any new wireless standard or technology are to be validated. 

For next generation wireless research, a viable prototyping option has emerged known as software-defined radio (SDR)~\cite{sdr_Mitola}. SDR enables researchers to rapidly prototype a system. Researchers at Stanford~\cite{Bharadia2013}~\cite{hong2014}, Rice~\cite{Duarte2012}\cite{Duarte2014}, and Princeton~\cite{aryafar2012} have implemented various testbeds to build in-band full-duplex radios using combined radio frequency (RF) antennas and SDR platform~\cite{Hong_FD_2015}. As shown in Fig.~\ref{fig:Exp_Setup}(a), a real-time full duplex LTE system was also demonstrated at IEEE Globecom  in Austin, TX, USA in December 2014.\footnote{The demo video is available at http://www.cbchae.org/.} 
The categorized comparison of implementation characteristics by each research group is summarized in Fig. \ref{fig:Exp_Setup}(b). The two sections that follow elaborate on how to solve the key challenge and implement real-time full duplex radios.


\section{Prototype Settings: System Specifications $\&$ Hardware Architecture} \label{Settings}
The demonstrated full duplex prototype~\cite{globecom2014} is based on the LTE downlink standard \cite{sesia2009lte} with the following system specifications: a transmission bandwidth of 20 MHz, 30.72 MHz sampling rate, 15 kHz subcarrier spacing, 2048 fast Fourier transform (FFT) size, and variable 4/16/64 quadrature amplitude modulation (QAM). The prototype is implemented, as shown in Fig. \ref{fig:Exp_Setup}(a), using LabVIEW system design software and state-of-the-art PXIe SDR platform, where two full duplex nodes consist of the following four main components. 

 $\bullet$	{\bf Dual-Polarization Full Duplex RF Antenna} Dual-polarization slot antenna with high cross-polarization discrimination (XPD) in all directions \cite{oh2014}.

 $\bullet$ {\bf PXIe-8133} Real-time (RT) controller equipped with a 1.73 GHz quad-core Intel Core i7-820 processor and 8 GB of dual-channel 1333 MHz DDR3 random access memory (RAM)~\cite{ni8133}. 

 $\bullet$	{\bf NI 5791R} 100 MHz bandwidth baseband transceiver module equipped with dual 130 MS/s analog-to-digital converter (ADC) with 14-bit accuracy, and dual 130 MS/s digital-to-analog converter (DAC) with 16-bit accuracy\cite{ni5791}.

 $\bullet$	{\bf PXIe-7965R} Field-programmable gate array (FPGA) module equipped with a Virtex-5 SX95T FPGA optimized for digital signal processing, 512 MB of onboard RAM, and 16 direct memory access (DMA) channels for high-speed data streaming at more than 800 MB/s\cite{ni7965}.
    \begin{figure*}
    \centering
    \includegraphics[width = 5.5in]{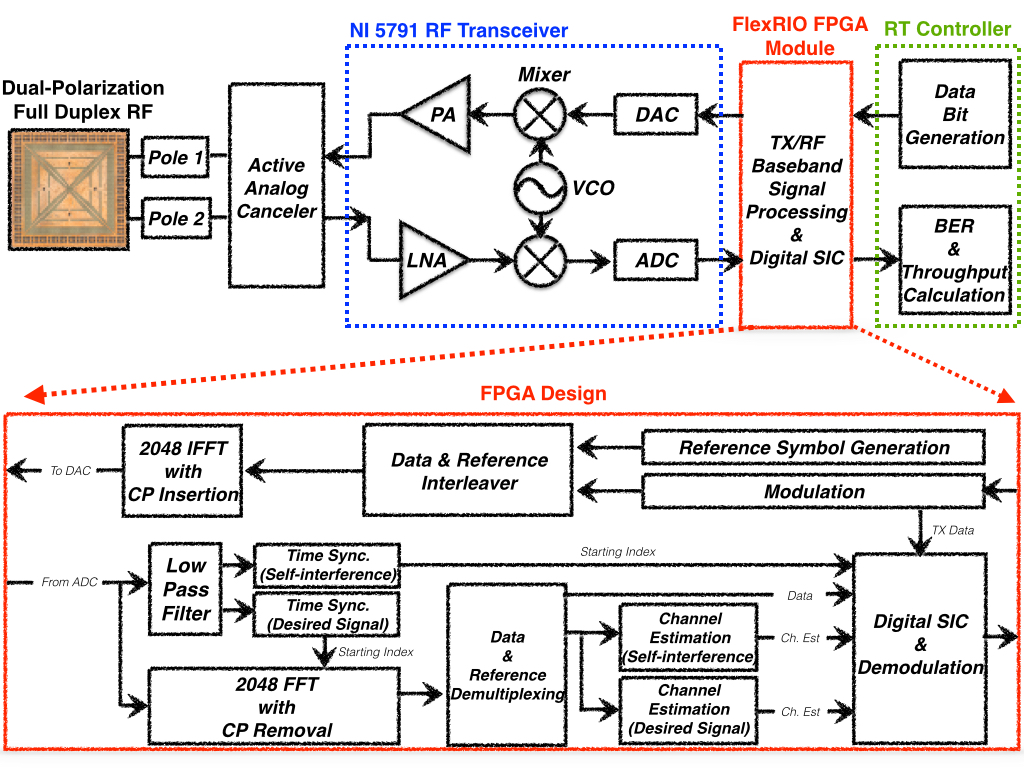}
    \caption{Block diagram of the proposed full duplex radio architecture. }              
    \label{fig:BD_FD}
    \end{figure*}
In addition, all these modules, except for the analog cancellation part including a dual-polarization full duplex RF antenna, sit in the NI PXIe-1075 chassis. The chassis plays a role in  data aggregation with both FPGA processors and a RT controller for real-time signal processing. As explained above, for transmitting and receiving simultaneously, the NI 5791R transceiver includes both transmit (Tx) and receive (Rx) ports connected with DAC and ADC, respectively.

As can be seen in Fig. \ref{fig:Exp_Setup}(a), we constructed a link for full duplex radios in an exhibition hall (a severe channel environment), where a great crowd of people was present, as well as in an indoor open space environment. The distance between full duplex communicating nodes was about 1.2~m. Note that in fact much longer ranges are possible. In this demo/experiment,  one transceiver is connected with a dual-polarization full duplex RF antenna (in the white box in Fig.~\ref{fig:Exp_Setup}(a)), and the other is connected with an omni-antenna for simplicity. In other words, the transceiver connected with the omni-antenna only transmits an uplink signal. We then observe results at the transceiver equipped with a full duplex RF antenna, where both the Tx and Rx ports are connected.



\section{Proposed Full Duplex System}
In this section, we elaborate, in processing order, on our design blocks for the real-time full duplex LTE system, from transmission to reception and self-interference cancellation. 
The block diagram of our full duplex radio architecture is illustrated in Fig. \ref{fig:BD_FD}. 

    \begin{figure*}
    \centering
    \includegraphics[width = 7in]{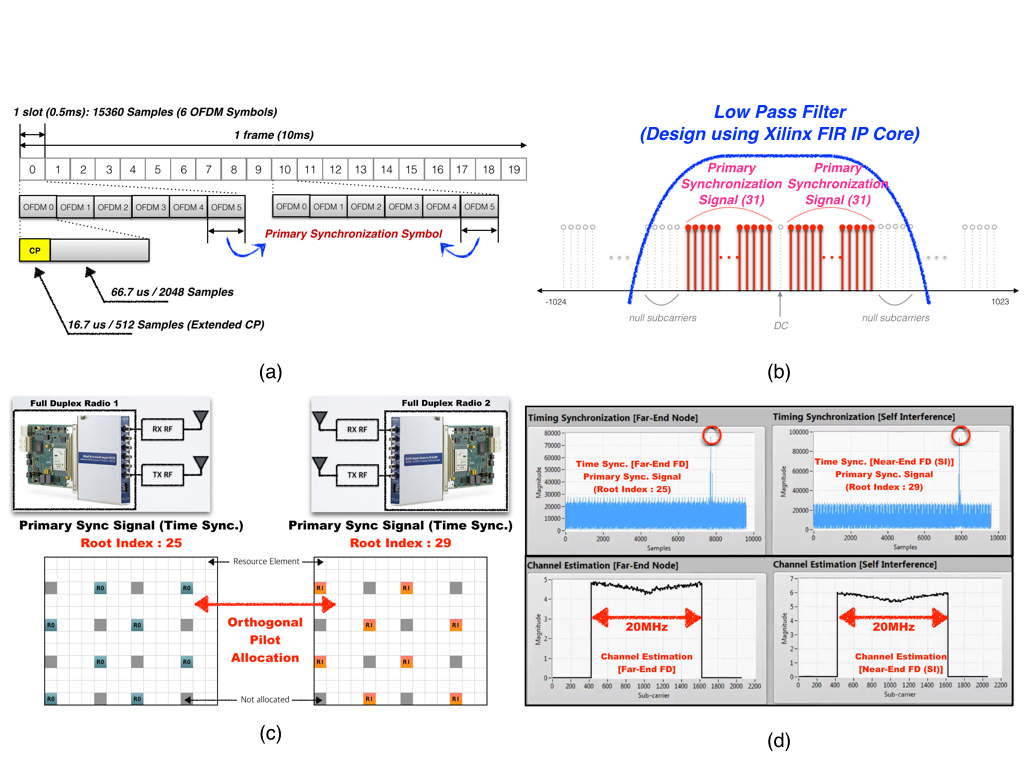}
    \caption{(a) Frame structure of our prototype (b) arrangement of primary synchronization symbol in frequency domain (c) reference symbol patterns for a full duplex link (d) measurement of synchronization and channel estimation in front panel of our prototype.}              
    \label{fig:Transmission}
    \end{figure*}

 \subsection{Transmission} 
As illustrated in Fig. \ref{fig:Transmission}(a), we follow the frame structure of the LTE downlink with a frame duration of 10~ms for transmission. Each frame is divided into 20 slots, each being 0.5 ms in duration. Each slot contains 6 orthogonal frequency division multiplexing (OFDM) symbols with 512 cyclic prefix (CP) length (extended mode).
The data bit is generated on the PXIe-8133 RT controller. After the modulation block, the data symbols are interleaved with reference symbols (RSs) stored in a look-up table. An array of interleaved symbols is padded with zeros to form an array of 2048 samples. The 2048 samples are passed through a 2048-point inverse FFT (IFFT) block transforming the frequency domain samples into the time domain. The 2048  IFFT with 512 CP insertion block is executed on the PXIe-7965R FPGA module. To operate the discrete Fourier transform (DFT), it uses Xilinx {\it fast Fourier transform}  intellectual property (IP) core.

 \subsection{Analog Self-interference Cancellation}
Conventional approaches to deal with self-interference as passive analog cancellation are 1) isolation between Tx and Rx signals~\cite{Bharadia2013}~\cite{hong2014}, 2) antenna separation between the Tx and Rx antennas~\cite{Duarte2012}\cite{Duarte2014}, or 3) signal inversion with a $\pi$-phase shifter~\cite{aryafar2012}. Although these strategies have been extensively studied and adapted to full duplex radios as a good solution, We focus on a simpler, more compact strategy that provides outstanding self-interference cancellation performance. Furthermore, we discuss a solution that provides more robustness to environmental effects, such as the Doppler effect and multi-path, to achieve stable analog cancellation performance.

For analog self-interference cancellation, we introduce a novel RF antenna. Our approach is based on a dual-polarization antenna with a high XPD characteristic. XPD is defined as the ratio of the co-polarized average received power to the cross-polarized average received power. It represents, in other words, the ability to maintain radiated or received polarization purity between horizontally and vertically polarized signals. As shown in Fig. \ref{fig:Exp_Setup}(a), the proposed RF unit is a compact antenna with two poles. One pole is used as a radiated Tx output; the other is used as a received Rx input in a full duplex radio. XPD is an important characteristic, particularly in full duplex systems, where cross-talk between Tx and Rx ports can curb the system's throughput performance. Since XPD has a relationship to inter-port isolation, the dual-polarization antenna with high XPD is, in full duplex systems, an excellent solution. We find that the dual-polarization antenna itself achieves 42 dB of isolation. Active analog cancellation provides an additional cancellation gain up to 18 dB by tuning the attenuation, phase shift, and delay parameters, i.e., totally 60 dB by analog cancellation.

 \subsection{Digital Self-interference Cancellation}
The goal of digital self-interference cancellation is to suppress, after canceling self-interference in the analog domain, any residual self-interference. Digital self-interference cancellation consists of rebuilding self-interference and subtracting it from the received signal. A key parameter to consider in the real-time digital self-interference canceler is the guaranteed throughput of digital data, in a given time, between transmitting and receiving streams in a node. Unlike unidirectional communications that uses a radio only for transmission or reception, a digital self-interference canceler of full duplex node requires, without a bottleneck, high speed computation/data throughput between transmit and receive elements. To perform rebuilding self-interference and subtracting it from the received signal in the real-time, we use handshake protocols, shift registers, shared registers with scheduled access, and dedicated first-in first-out (FIFO) buffers.

In \cite{choi2013}, the authors implemented a full duplex solution, which needs no additional synchronization or channel estimation for self-interference in the digital domain. Our prototype, however, focuses on an independent system that operates in real-time to maximize cancellation performance in the analog/digital domain, respectively. Further, additional synchronization and estimation is exploited to reduce the complexity in our digital self-interference canceler. 

At the moment of decoding the desired symbol, it is critical to know the perfect timing between self-interference and the received symbol in full duplex mode. Thus, key issues include designing synchronization and channel estimation strategies for residual self-interference as well as for a desired link. We produce a process for implementing a digital self-interference canceler from synchronization and channel estimation. In order to operate a real-time digital self-interference canceler with high performance, we focus on FPGA implementation using LabVIEW system design software and PXIe SDR platform.

     \begin{figure*}
    \centering
    \includegraphics[width = 6.3in]{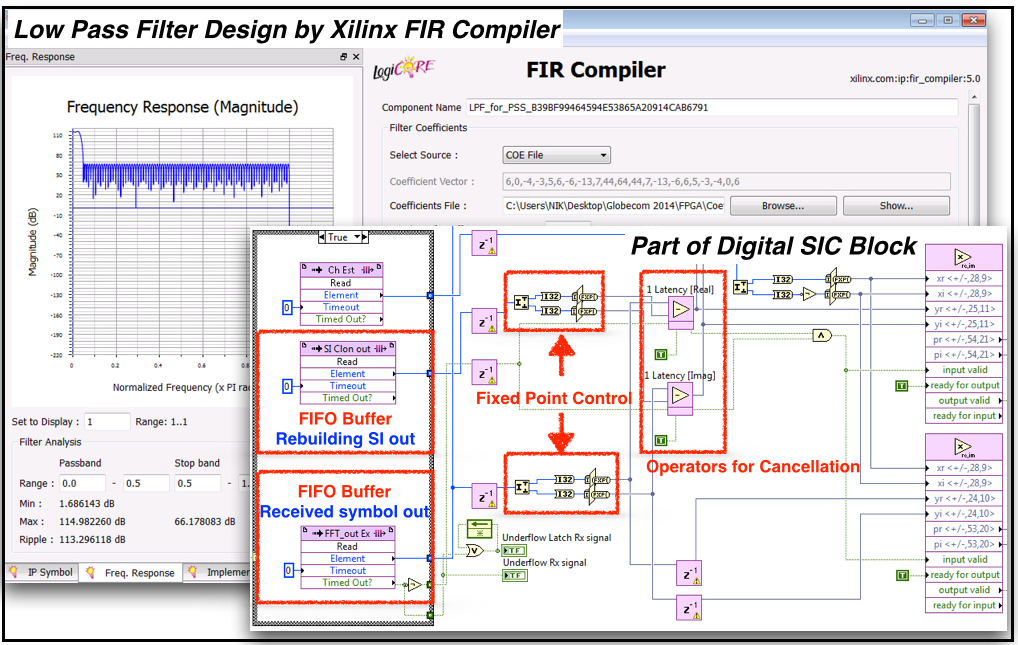}
    \caption{A part of source code for a digital self-interference canceller that operates in real-time on the software-defined radio platform.  }              
    \label{fig:LV}
    \end{figure*}
 
$\bullet$ {\bf Synchronization:} Synchronization is one of the key blocks in real-time full duplex radios. Under synchronization for full duplex, there are two operations:  synchronization for decoding the desired symbol and rebuilding self-interference. In the synchronization block for decoding the desired symbol, we estimate time offset by random propagation delays and sampling clock offsets between two full duplex radios. In the synchronization block for rebuilding self-interference, we estimate the time offset between Tx port and Rx port of a full duplex radio.

To facilitate timing synchronization, the LTE downlink standard specifies a primary synchronization signal (PSS). Accordingly, the receiver can successfully perform timing synchronization in half-duplex mode. Note, however, that we need to keep performing synchronization for the self-interference signal as well as for the desired signal. Thus, we use a property, where the ZC sequence with a different root index is orthogonal to each other. The PSS is modulated by a ZC sequence given as, $P\left[ k \right] = {e^{ - j\frac{\pi }{N}uk\left( {k + 1} \right)}}, - 31 \le k \le  - 1$, and $P\left[ k \right] = {e^{ - j\frac{\pi }{N}u\left( {k + 1} \right)\left( {k + 2} \right)}}, 1 \le k \le 31$, where $k$ is the subcarrier index, $u$ is the root index, and $N$ it the sequence length ($N$ = 63). We use a different root index relatively prime to $N$ for the PSS of each full duplex radio, i.e, $u_1$ = 25, $u_2$ = 29. These symbols are located on the 62 subcarriers, symmetrically arranged around the DC-carrier in the last OFDM symbol of the first and eleventh slots of each frame as shown in Fig. \ref{fig:Transmission}(a) and (b). As the duration of a frame is 10 ms, the PSS is therefore transmitted after every 5 ms time intervals or once per half-frame. 

To calculate the correlation between the ideal sequence and the estimated PSS signal, it is necessary to extract the PSS subcarrier from the received signal. For this reason, we design a low-pass filter (LPF) using Xilinx's finite impulse response (FIR) IP core, as shown in Fig. \ref{fig:Transmission}(b). The designed LPF has a cut-off frequency of 1.4 MHz, a stop-band attenuation of 50~dB, and a pass-band ripple of 0.1 dB. After the received signal samples are passed through the LPF, each synchronization block for decoding the desired symbol and rebuilding the self-interference is executed to calculate, independently, the correlation with its own PSS. As a result, a maximum peak is detected at the sample index of the first sample of the OFDM symbol following the PSS symbol, as illustrated in Fig. \ref{fig:Transmission}(d). A starting index of the desired signal is delivered into FFT block, and a starting index of self-interference signal is delivered into the digital cancellation block.

    \begin{figure*}
    \centering
    \includegraphics[width = 6.3in]{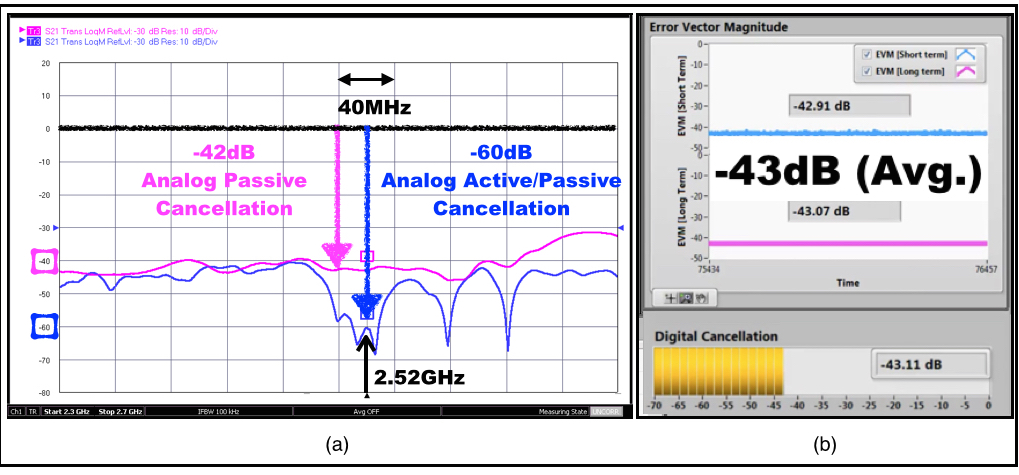}
    \caption{Measurement results of (a) analog self-interference cancellation and (b) digital self-interference cancellation measurement.}                
    \label{fig:AD_SIC}
    \end{figure*}

$\bullet$ {\bf Channel Estimation:} Channel estimation also has, for full duplex, two operations: estimations for 1) the channels between two full duplex nodes and 2) the channel between the Tx and Rx ports in its own full duplex node. The former operates to decode the desired symbol after digital self-interference cancellation, while the latter operates to rebuild the self-interference using the known Tx data. In order to handle the two operations simultaneously, we design RS patterns that are orthogonal between two full duplex nodes, exploiting the pattern of cell-specific reference signals for multiple antenna port. The RS patterns are shown in Fig.~\ref{fig:Transmission}(c). 

Both channel estimation blocks have two steps in common: 1) RS extraction and 2) interpolation in order. After the received samples are passed through the FFT block, the RS subcarriers of each channel are extracted from an OFDM symbol in a data and reference demultiplexing block. A  channel coefficient of each RS subcarrier is calculated using original RSs stored in block memory. To estimate the channel coefficients of RS subcarriers, a least-square method is exploited. After passing through the data and reference demultiplexing block, the channel estimates of the RS subcarriers are split into two groups. One is for the channel estimation between nodes; the other is for the channel estimation between antenna ports. In each channel estimation block, we implemented a linear interpolator using Xilinx's FIR IP core. The linear interpolator in each block estimates the channel coefficients of data subcarriers as well as RS subcarriers. 
In Fig. \ref{fig:Transmission}(d), the bottom left and right figures are screen shots of the instantaneous channel estimation result between nodes and ports, respectively, in frequency domain.

$\bullet$ {\bf Digital Cancellation:} In most research on full duplex implementation, digital self-interference cancellation is performed in the time domain. To carry out digital cancellation in the time domain, an additional IFFT block is needed for rebuilding self-interference. As mentioned above, in the real-time digital self-interference canceler, it is critical to operate the process for rebuilding self-interference and canceling it out without a bottleneck. For lower computational complexity and faster rebuilding self-interference, we execute it in the frequency domain. One might argue that if the SI is not overlapped coherently with the received signal, it would be difficult to cancel the SI in the frequency domain. If the difference, however, in the received time of the SI and the desired signal is less than the CP length, which is common in practice, there is no problem in canceling the SI after FFT. We will also investigate this issue in our future work. 

Digital cancellation utilizes the baseband samples of the transmitted signal to rebuild self-interference in the digital domain and subtracts them from the received samples. Note that we know the baseband samples of the transmitted signal from its own node. Self-interference can be rebuilt in the digital domain using the baseband samples of the transmitted signal and the channel estimates between the ports of its own node. As mentioned above, we should know which self-interference (a sample index) is mixed in the received sample at the moment of decoding the desired symbol. Accordingly, we include a counter in the digital cancellation block. As soon as the starting index of the self-interference signal arrives in the digital cancellation block from the synchronization block to rebuild self-interference, the counter operates to choose a rebuilt digital sample for subtraction processing. After digital cancellation, a zero-forcing channel equalizer operates to decode the desired symbols. Illustrated in Fig. \ref{fig:LV} is a part of the source code for the digital self-interference canceler operated in the frequency domain.
  
\section{Prototype Test Results}
Using the real-time full duplex LTE prototype as described in the previous section, we measure the level of analog and digital self-interference and calculate the bit error rate (BER) and throughput performance. In this prototype, the carrier frequency is the 2.52 GHz in LTE bands. As shown in Fig.~\ref{fig:AD_SIC}(a), we find that the dual-polarization antenna provides about 42 dB of isolation from our experiments, i.e., the self-interference that is leaking to the Rx port is reduced by about 42 dB. Moreover, by tuning the attenuation, phase shift, and delay parameters, we achieve 60 dB of analog self-interference cancellation with analog active cancellation. In the digital domain, we calculate error vector magnitude (EVM) for self-interference to measure the average level of digital self-interference cancellation. As a result, we achieve 43 dB of self-interference cancellation in the digital domain as shown in Fig. \ref{fig:AD_SIC}(b). 

\begin{figure}
    \centering
    \includegraphics[width = 3.5in]{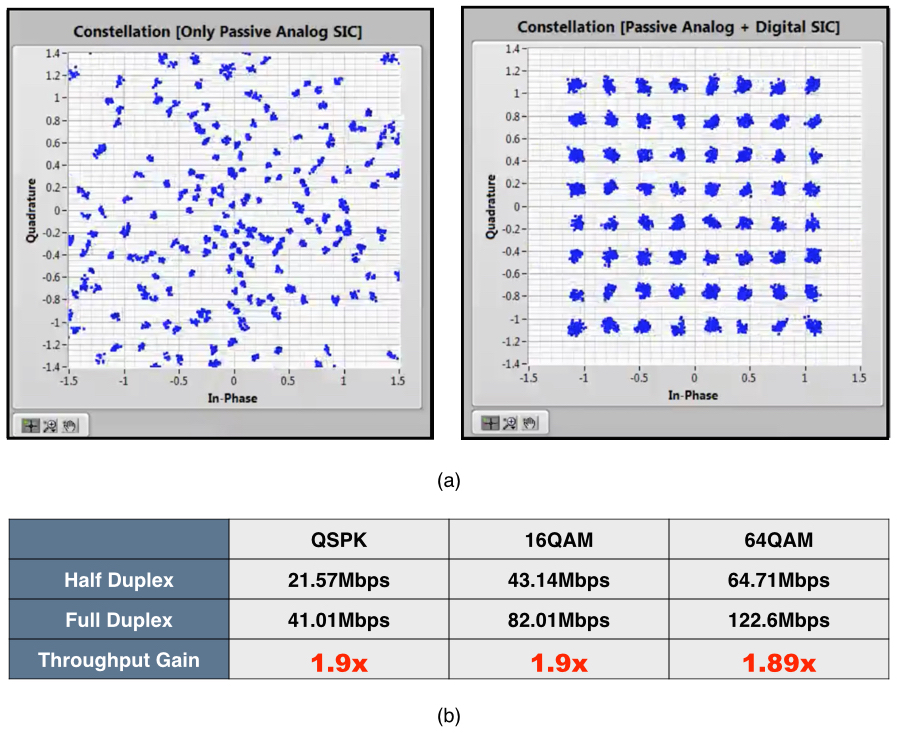}
    \caption{(a) Constellation result with only analog passive self-interference cancellation (left) constellation result with analog and digital self-interference cancellation (right) (b) throughput performances for each constellation.}              
    \label{fig:Cons_TP}
    \end{figure}

In order to compare the throughput improvement, we also implemented the LTE-FDD prototype. Our design is based upon the LTE downlink standard with system specifications that include a transmission bandwidth of 10 MHz, 15.36 MHz sampling rate, 15 kHz subcarrier spacing, 1024 FFT size, 256 CP length, and variable 4/16/64 QAM.
Fig.~\ref{fig:Cons_TP}(a) shows the constellation, taken during an over-the-air test of the full duplex communications link. One full duplex radio transmits a 4~QAM modulated signal as the downlink, and receives a 64 QAM modulated signal as the uplink. As a result, the goal of this full duplex radio is to decode 64 QAM, the desired symbol after perfectly canceling out the 4~QAM symbol as self-interference. In Fig. \ref{fig:Cons_TP}(a), the left constellation shows that, with only analog cancellation, self-interference is not perfectly cancelled out, while the right constellation shows that with both analog and digital cancellation self-interference is perfectly cancelled out.

As shown in Fig. \ref{fig:Cons_TP}(b), our prototype provides high throughput performance in real-time. It delivers a throughput increase of 1.9x on the 4, 16~QAM and 1.89x on the 64~QAM compared to the conventional half duplex mode.\footnote{Half duplex means out-of-band full duplex, i.e., FDD throughout this paper.}

\section{Research Challenges}
Notwithstanding our focus on designing more practical full duplex radios, several research challenges remain before the most viable solution for next generation communication systems is achieved. 

\subsection{Hardware Impairments}
The performance of a full duplex system depends heavily on hardware impairments: amplifier non-linearity, gain/phase offset, I/Q imbalance, quantization effects, and timing jitter. For example, a nonlinearly amplified OFDM signal occurs intermodulation distortion (IMD), which the amplitude modulation of signals containing two or more different frequencies in a system. The IMD raises the noise floor and causes inter-carrier interference, which induce performance degradation of the full duplex system. Because most analog components in the system have the non-linearity property, the cancellation of all non-linear components as well as linear components is a significant burden on a real-time system. Thus, some pre-processing strategies such as pre-distortion for reducing hardware impairments represent an interesting research topic.

\subsection{Joint PHY/MAC Prototyping}
Most implementations of full duplex radios have mainly focused on the physical layer design, which enables bi-directional communications between a single pair link. There exists apparent limitations in translating the performance gains obtained from the demonstration of a single pair link into network performance.
Transmissions on full duplex mode create potential interference outside the full duplex link. This calls for the prototyping of media access control (MAC) layer protocols, including discovering and exploiting full duplex opportunities in a distributed manner. Another interesting area for future work is the joint PHY/MAC approach for prototyping. 

\subsection{Full Duplex System with OFDM and SC-FDMA}
Since single carrier frequency division multiple access (SC-FDMA) has a peak-to-average power
ratio (PAPR) lower than that of OFDMA, it is used for the uplink multiple access scheme in the LTE of cellular systems. Most implementation studies of full duplex, however, deal with only OFDM frame structures. There are many potential challenges in asymmetric uplink/downlink frame structures in LTE.

\subsection{Comparison with LTE-TDD}

LTE-TDD (time division duplexing), generally, has been known to have many benefits, such as low latency, spectrum flexibility, uplink/downlink flexibility, and lower cost per bit. To discuss the various performance characteristics such as latency, throughput, power consumption, and flexibility between TDD and full duplex, we believe that it is worth comparing full duplex prototype with a comparable LTE-TDD prototype.

\subsection{Novel Solution for RF/Analog Cancellation}
RF/Analog cancellation plays a critical part in attenuating high-powered self-interference sufficiently such that Rx saturation and dynamic range are not an issue when operating digital cancelation. We showed the analog cancellation based on the dual polarization would be a good option. It does have, however, two main weaknesses. First, it struggles to perform active analog cancellation in real-time. While high XPD makes the passive analog cancellation level high, to estimate the coefficients for active cancellation is difficult. Second, channel reciprocity that can simplify the link overhead may not be assumed for the polarization antennas. The novel solutions for these issues will be an interesting research topic.
 
\section{Conclusion}
Full duplex radio technologies could be a major contributor to increasing spectrum efficiency in areas of explosive traffic demand where there are limited radio resources. To validate the feasibility and commercial viability of any new wireless standard or technology like full duplex radio, SDR-based prototyping is imperative. We prototyped a design that combines dual-polarization full duplex RF and the digital self-interference canceler that operates in real-time on the SDR platform. We focused on a more practical prototype that exhibited outstanding self-interference cancellation performance. The main portion of this article is dedicated to presenting the design, implementation, and evaluation of a real-time full duplex LTE system, a candidate for next generation wireless communication systems. We expect our prototype design to provide worthwhile insights into developing the most viable solution for future wireless communication systems.


\ifCLASSOPTIONcaptionsoff
  \newpage
\fi




%

\renewcommand{\baselinestretch}{1.0}
\bibliographystyle{IEEEtran}
\bibliography{reference_CommMag14} 

\end{document}